\pdfoutput=1

\documentclass[10pt,a4paper,pra,aps,showpacs,showkeys,twocolumn]{revtex4-1}
\usepackage{hyperref}
\usepackage{graphicx}
\usepackage{subfigure}
\usepackage{amsmath,amssymb}

\newcommand{\eg}{\emph{e.g.\/}}
\newcommand{\ie}{\emph{i.e.\/}}
\newcommand{\etal}{\emph{et al.}}
\DeclareMathOperator{\tr}{tr}

\newcommand{\ket}[1]{\ensuremath{|#1\rangle}}
\newcommand{\bra}[1]{\ensuremath{\langle#1|}}
\newcommand{\ketbra}[2]{\ensuremath{\ket{#1}\bra{#2}}}
\newcommand{\braket}[2]{\ensuremath{\langle{#1}|{#2}\rangle}}
\newcommand{\1}{{\rm 1\hspace{-0.9mm}l}}
\newcommand{\Id}{\1}

\begin{document}

\title{Quantum walks with memory on cycles}

\author{Michael Mc Gettrick}
\affiliation{De Br\'un Centre for Computational Algebra, School of Mathematics,
National University of Ireland, \\ University Road, Galway, Ireland}

\author{Jaros{\l}aw Adam Miszczak}
\affiliation{Institute of Theoretical and Applied Informatics, Polish Academy
of Sciences,\\ Ba{\l}tycka 5, 44-100 Gliwice, Poland}

\begin{abstract}
We study the model of quantum walks on cycles enriched by the addition of 1-step
memory. We provide a formula for the probability distribution and the
time-averaged limiting probability distribution of the introduced quantum walk.
Using the obtained results, we discuss the properties of the introduced model
and the difference in comparison to the memoryless model.
\keywords{quantum walks, Markov processes, limiting distribution}
\pacs{03.67.-a, 05.40.Fb, 02.50.Ga}
\end{abstract}

\date{04/09/2013 (v.~0.72)}

\maketitle 

\section{Introduction}
During the last few years a considerable research effort has been made to
develop new algorithms based on the rules of quantum mechanics. Among the
methods used to achieve this goal, quantum walks, a quantum counterpart of
random walks, provide one of the most promising and successful approaches.

Classical random walks can be applied to solve many computational problems. They
are used, for example, to find spanning trees and shortest paths in graphs, to
find the convex hull of a set of points or to provide a sampling-based volume
estimation~\cite{reitzner12walks}. Today a huge research effort is devoted to
applying random walks in different areas of science. Classical random walks find
their application in a plethora of areas. This has motivated big interest in
using a similar model for developing algorithms which could harness the
possibilities offered by quantum machines. 

Quantum walks are counterparts of classical random walks governed by the rules
of quantum mechanics~\cite{venegas08synthesis,vanegas12review,reitzner12walks}
and provide a promising method for developing new quantum algorithms. Among the
applications of quantum walks one can point out: solving the element
distinctness \cite{ambainis07distinctness} and subset finding
\cite{childs05subset} problems, spatial search~\cite{childs04spatial}, triangle
finding~\cite{magniez05quantum} and verifying matrix
products~\cite{buhrman06quantum}. The survey of quantum algorithms based on
quantum walks is presented in~\cite{ambainis03algorithmic}.

The influence of memory on the behavior of quantum walks has been considered by
Flitney \etal~\cite{flitney04quantum} and Brun \etal~\cite{brun03quantum}.
In~\cite{mcgettrick10one} Mc Gettrick proposed a model of one-dimensional
quantum walk on line with one-step memory and studied the limiting probability
distribution for this model. This work was developed by Konno and Machida in
\cite{konno10limit}. More recently Rohde \etal\ considered a quantum walk with
memory constructed using recycled coins and applied numerical experiments to
study its properties. Moreover, an experimental proposal for implementing a
quantum walk with memory using linear optics has also been considered
in~\cite{rohde12quantum}.

In this paper we introduce and study the model of quantum walks on cycle
\cite{aharonov01walks,bednarska03walks} enriched by the addition of 1-step
memory~\cite{mcgettrick10one}. Our main contribution is the calculation of the
probability of finding the particle at each position after given number of
steps, and the limiting probability distribution for the introduced model. We
also point out the differences between quantum walks on cycles with and without
memory.

This paper is organized as follows.
In Section~\ref{sec:cycle-and-memory} we introduce the model of a quantum walk
on cycle with one-step memory.
In Section~\ref{sec:fourier-analysis} we analyze the introduced model using
Fourier transform method and
we discuss the behavior of the time-averaged
limiting probability distribution of the discussed model.
Finally, in Section~\ref{sec:final-remarks} we summarize the obtained results
and provide some concluding remarks.

\section{The model}\label{sec:cycle-and-memory}
In the model discussed in~\cite{bednarska03walks} the space used by a quantum
walk is composed of two parts -- 1-qubit coin and $d$-dimensional state space,
\ie\ $\mathcal{H}=\mathcal{H}_2\otimes \mathcal{H}_d$. The shift operator 
in this case is defined as
\begin{equation}
S_0 = \sum_{c=0}^1\sum_{v=0}^{d-1} \ketbra{c}{c}\otimes \ketbra{v+2c-1\bmod d}{v}.
\end{equation}

Here we adopt this model and extend it with an additional register, referred to
as memory register, which stores the history of a walk. 

For a quantum walk with one step memory one needs a single qubit to store the
history. In this case we use a Hilbert space of the form
$\mathcal{H}=\mathcal{H}^{c}_2\otimes\mathcal{H}^{\mathrm{m}}_2\otimes\mathcal{H}_d$,
respectively for a coin, memory and a position.

As in the case of a memoryless walk, the first register is the coin register and
the third register is used to encode the position of the particle. The second
register stores the history of the walk. The history is encoded as direction
from which the particle was moved in the previous move. If this register is in
the state $\ket{0}$, the previous position of the particle was $n+1$. If this
register is in the state $\ket{1}$, the previous position of the particle was
$n-1$. The coin register indicates if the walk should continue in the previously
chosen direction (transmission in state $\ket{0}$) or change the direction
(reflection in state $\ket{1}$).

Taking into account the above, we define a shift operator for a quantum walk
with a 1-step memory on cycle with $d$ nodes as
\begin{equation}\label{eqn:shift-cycle-memory}
\begin{split}
S_1  = \sum_{n=0}^{d-1}&\Big(
\ketbra{0}{0}\otimes \ketbra{0,n-1(\bmod d)}{0,n}\\
+&\ketbra{0}{0}\otimes \ketbra{1,n+1(\bmod d)}{1,n}\\
+&\ketbra{1}{1}\otimes \ketbra{1,n+1(\bmod d)}{0,n}\\
+&\ketbra{1}{1}\otimes \ketbra{0,n-1(\bmod d)}{1,n}
\Big),
\end{split}
\end{equation}
or in a more consistent form as
\begin{equation}
S_1 = \sum_{n,m,c}\ketbra{c}{c}\otimes \ketbra{h_{m,c},n+2 h_{m,c}-1(\bmod d)}{m,n},
\end{equation}
where $h_{m,c} = m+c (\bmod 2)$ represents a history dependence of the walk.

The walk operator is defined as toss-a-coin and make-a-move combination, \ie\
\begin{equation}
W_1 = S_1(C\otimes \Id_2\otimes \Id_d),
\end{equation}
where $C$ is a coin matrix, \eg\ Hadamard matrix $H$
\begin{equation}
H=\frac{1}{\sqrt{2}}\left(
\begin{matrix}
1 & 1\\
1 & -1
\end{matrix}
\right)
\end{equation}
or any matrix $C\in\mathrm{SU}(2)$.

The walk starts in some initial state $\ket{\phi_0}$. After each step the state
is changed according to the formula
\begin{equation}
\ket{\phi_n} = W_1^n\ket{\phi_{0}}
\end{equation}
or as a recursive relation
\begin{equation}
\ket{\phi_{n+1}} = W_1\ket{\phi_{n}}.
\end{equation}

The probability of finding a particle at position $v$ after $n$ steps is
obtained after averaging over the coin and the memory registers
\begin{equation}
p(v,n) = \sum_{c,m}|\braket{c,m,v}{\phi_n}|^2,
\end{equation}
or, in other words, by tracing out over the memory and the coin subspaces
\begin{equation}
p(v,n) = \tr_{c,m}\ketbra{\phi_n}{\phi_n},
\end{equation}
where $\tr_{c,m}$ denotes the operation of tracing out with respect to the coin
and the memory subspaces.

In \cite{aharonov01walks} it was shown that $p(v,n)$ is quasi-periodic for
memoryless walks on cycles and it was suggested to consider quantity
\begin{equation}
\bar{p}(n,t) = \frac{1}{t}\sum_{s=1}^t p(n,s),
\end{equation}
which converges with $t\rightarrow\infty$ to the limiting distribution $p(v)$.
As the parameter $n$ in this formula corresponds to the time required to perform
$n$ steps, we refer to so defined $\bar{p}(v)$ as time-averaged limiting
distribution.

\section{Probability distribution}\label{sec:fourier-analysis}
In what follows we evaluate the probability distribution of finding the particle
at each node of the cycle. We consider the Hadamard walk only. Thus the walk
operator is given as
\begin{equation}\label{eqn:walk-op}
W=S(H\otimes \1_{2 d}).
\end{equation}

The second factor in the Eq.~(\ref{eqn:walk-op}) can be written in the matrix
notation as
\begin{equation}
H\otimes \1_{2 d} = 
\frac{1}{\sqrt{2}}\left(
\begin{matrix}
\1_{2d} & \1_{2d} \\
\1_{2d} & - \1_{2d} \\
\end{matrix}
\right).
\end{equation}

\subsection{Fourier analysis}
In order to obtain expression for the probability distribution on the cycle we
use Fourier analysis method~\cite{nayak00line,grimmett04waek,konno10limit}. 

Below we calculate amplitudes for a walk on cycle with a 1-step memory with the
Hadamard matrix acting on the coin register. In this case we represent the
vectors of amplitudes as
\begin{equation}
\Phi(n,t) = \left(
\begin{matrix}
\braket{0,0,n}{\psi_t}\\
\braket{0,1,n}{\psi_t}\\
\braket{1,0,n}{\psi_t}\\
\braket{1,1,n}{\psi_t}
\end{matrix}
\right).
\end{equation}

The shift operator is defined as in Eq.~(\ref{eqn:shift-cycle-memory}). In
this case the interesting part of the shift operator reads
\begin{equation}
\begin{split}
S_1(n)&= \ketbra{0}{0}\otimes\ketbra{0,n}{0,n+1}+\ketbra{0}{0}\otimes\ketbra{1,n}{1,n-1}\\
&+\ketbra{1}{1}\otimes\ketbra{0,n}{1,n+1}+\ketbra{1}{1}\otimes\ketbra{1,n}{0,n-1}
\end{split}.
\end{equation}
After one step of time evolution we have
\begin{eqnarray*}
\Phi(n,t+1) = 
\left(
\begin{matrix}
\bra{0,0,n+1}(H\otimes\1_2\otimes\1_d)\ket{\psi_t}\\
\bra{0,1,n-1}(H\otimes\1_2\otimes\1_d)\ket{\psi_t}\\
\bra{1,1,n+1}(H\otimes\1_2\otimes\1_d)\ket{\psi_t}\\
\bra{1,0,n-1}(H\otimes\1_2\otimes\1_d)\ket{\psi_t}
\end{matrix}
\right).
\end{eqnarray*}
Evaluating the action of the Hadamard gate on the coin register one gets
\begin{eqnarray*}
\Phi(n,t+1) &&= \frac{1}{\sqrt{2}}\left(
\begin{smallmatrix}
\braket{0,0,n+1}{\psi_t} + \braket{1,0,n+1}{\psi_t}\\
0\\
\braket{0,1,n+1}{\psi_t} - \braket{1,1,n+1}{\psi_t}\\
0
\end{smallmatrix}
\right) \\
&&+ 
\frac{1}{\sqrt{2}}\left(
\begin{smallmatrix}
0\\
\braket{0,1,n-1}{\psi_t} + \braket{1,1,n-1}{\psi_t}\\
0\\
\braket{0,0,n-1}{\psi_t} - \braket{1,0,n-1}{\psi_t}
\end{smallmatrix}
\right).
\end{eqnarray*}

Rewriting the above expression using $\Phi(n+1,t)$ and $\Phi(n-1,t)$ one gets
\begin{equation}
\Phi(n,t+1) = M_{-} \Phi(n+1,t) + M_{+} \Phi(n-1,t),
\end{equation}
where $M_{+}$ (\emph{advancing}) and $M_{-}$ (\emph{retarding}) matrices read
\begin{eqnarray}
M_{+} &=& \frac{1}{\sqrt{2}}
\left(
\begin{smallmatrix}
0 & 0 & 0 & 0 \\
0 & 1 & 0 & 1 \\
0 & 0 & 0 & 0 \\
1 & 0 & -1 & 0 
\end{smallmatrix}
\right),\\
M_{-} &=& \frac{1}{\sqrt{2}}
\left(
\begin{smallmatrix}
1 & 0 & 1 & 0 \\
0 & 0 & 0 & 0 \\
0 & 1 & 0 & -1 \\
0 & 0 & 0 & 0
\end{smallmatrix}
\right).
\end{eqnarray}

In order to obtain the expression for the amplitudes of the quantum walk with
memory on cycle we use the method introduced in~\cite{nayak00line} and represent
time evolution of the walk using the Fourier transform
\begin{eqnarray}
\tilde\Phi(k,t+1) &=& \sum_{n=0}^{d-1} e^{2 \pi i kn/d} \Phi(n,t+1)\\\nonumber
&=& \sum_{n=0}^{d-1} e^{2 \pi i kn/d} (M_{-} \Phi(n+1,t) + M_{+} \Phi(n-1,t)) \\\nonumber
&=&  \left( e^{-2 \pi i kn/d} M_{-}  + e^{2 \pi i kn/d} M_{+}\right) \tilde\Phi(k,t).
\end{eqnarray}
From the above we get a recursive relation for the time evolution in the Fourier
basis
\begin{equation}
\tilde\Phi(k,t) = M_k^t \tilde\Phi(k,0),
\end{equation}
where
\begin{equation}
M_k = \frac{1}{\sqrt{2}}\left(
\begin{array}{cccc}
 e^{-\frac{2 i k \pi }{d}} & 0 & e^{-\frac{2 i k \pi }{d}} & 0 \\
 0 & e^{\frac{2 i k \pi }{d}} & 0 & e^{\frac{2 i k \pi }{d}} \\
 0 & e^{-\frac{2 i k \pi }{d}} & 0 & -e^{-\frac{2 i k \pi }{d}} \\
 e^{\frac{2 i k \pi }{d}} & 0 & -e^{\frac{2 i k \pi }{d}} & 0
\end{array}
\right).
\end{equation}

Let us now denote $A=\exp(-2 \pi i k/d) $. Matrix $M_k$ has the following
eigenvalues
\begin{eqnarray*}
\lambda_1 &=& -1, \\
\lambda_2 &=& 1, \\
\lambda_3 &=& \frac{1+A^2+\sqrt{A^4-6 A^2+1}}{2 \sqrt{2} A},\\
\lambda_4 &=& \frac{1+A^2-\sqrt{A^4-6 A^2+1}}{2 \sqrt{2} A},
\end{eqnarray*}
with corresponding (orthonogal, but unnormalized) eigenvectors
\begin{eqnarray*}
v_1 &=& \left\{ 
\begin{smallmatrix}
-\frac{\sqrt{2}
 A^2}{2 A+\sqrt{2}}\\
-\frac{1}{\sqrt{2} A+1}\\
\frac{A \left(\sqrt{2} A+2\right)}{2 A+\sqrt{2}}\\
1
\end{smallmatrix}
\right\},
\quad
v_2 = \left\{
\begin{smallmatrix}
\frac{A^2}{\sqrt{2} A-1}\\
\frac{1}{\sqrt{2} A-1}\\
\frac{A \left(\sqrt{2} A-2\right)}{\sqrt{2}-2 A}\\
1
\end{smallmatrix}
\right\},\\ 
v_3 &=& \left\{
\begin{smallmatrix}
\frac{1}{2} \left(A^2-\sqrt{A^4-6 A^2+1}-1\right)\\
\frac{A^2+\sqrt{A^4-6 A^2+1}-1}{2 A^2}\\
-1\\
1
\end{smallmatrix}
\right\},\\
v_4 &=& \left\{
\begin{smallmatrix}
\frac{1}{2} \left(A^2+\sqrt{A^4-6 A^2+1}-1\right)\\
-\frac{-A^2+\sqrt{A^4-6 A^2+1}+1}{2 A^2}\\
-1\\
1
\end{smallmatrix}
\right\}.
\end{eqnarray*}


\subsection{Initial state and time evolution}
In order to evaluate the probability distribution we need to choose the initial
state of the walk. Again following~\cite{nayak00line} we start from the position
$0$ with the coin in the state $\ket{0}$. As in this case we have extra register
for storing memory, in the initial state the state vector reads
\begin{equation}\label{eqn:init-state}
\ket{\phi_0} = 
\frac{1}{\sqrt{2}}\left(
\begin{smallmatrix}
1\\ 1\\ 0\\ 0
\end{smallmatrix}
\right)\otimes\ket{0},
\end{equation}
which means that the memory register is in the superposition
$\frac{1}{\sqrt{2}}(\ket{0}+\ket{1})$ and the coin register is in the state
$\ket{0}$.The vector of amplitudes in this situation reads
\begin{equation}
\Phi(n,0) = \frac{1}{\sqrt{2}}
\left(
\begin{smallmatrix}
1\\1\\0\\0
\end{smallmatrix}
\right) \delta_{n0}.
\end{equation}
Thus in the Fourier basis we start from the state
\begin{equation}
\tilde\Phi(k,0) = 
\frac{1}{\sqrt{2}}\left(
\begin{smallmatrix}
1\\1\\0\\0
\end{smallmatrix}
\right),
\end{equation}
for any $k=0,\ldots,d-1$.

\subsection{Initial state decomposition}
Using the eigendecomposition of the matrix $M_k$ we can calculate the form of
the amplitudes in the Fourier basis after $t$ steps. After $t$ steps the vector
of the amplitudes reads
\begin{equation}
\tilde\Phi(k,t) = M_k^t \tilde\Phi(k,0) = 
 M_k^t\frac{1}{\sqrt{2}}\left(
\begin{smallmatrix}
1\\1\\0\\0
\end{smallmatrix}
\right)
\end{equation}
for any $k$.

Let us now write the initial state of the walk $\tilde\Phi(k,0)$ in the
$\{v_i\}_{i=1,\dots,4}$ basis as
\begin{equation}\label{eqn:fourier-init-state}
\tilde\Phi(k,0) = \sum_{i=1}^4 \alpha_i(k) v_i(k).
\end{equation}
where 
\begin{equation}\label{eqn:fourier-init-state-alphas}
\alpha_i(k) = (v_i(k), \tilde\Phi(k,0) )
\end{equation}
are components of
$\tilde\Phi(k,0)$ in $\{v_i\}_{i=1,\dots,4}$ basis. 
The evolution in the Fourier basis can be now written as
\begin{equation}
\tilde\Phi(k,t)  = M_k^t \sum_{i=1}^4 \alpha_i(k) v_i(k)
                 = \sum_{i=1}^4  \alpha_i(k) \lambda_i(k)^t v_i(k).
\end{equation}

The original components can be expressed by the Fourier-transformed vectors as
\begin{equation}
\begin{split}
\Phi(n,t) & =\frac{1}{d}\sum_{k=0}^{d-1}e^{2 \pi i k n /d}\tilde\Phi(k,t)\\
          & =\frac{1}{d}\sum_{k=0}^{d-1} \sum_{j=1}^4 e^{2 \pi i k n /d} \alpha_j(k) \lambda_j(k)^t v_j(k).
\end{split}
\end{equation}


Using the above, the probability of finding the particle at $n$-th node after
$t$ steps reads
\begin{equation}
\begin{split}\label{eqn:prob-n-t}
p(n,t) & = |\Phi(n,t)|^2\\
       & = \frac{1}{d^2} \sum_{k,m=0}^{d-1} \sum_{j,l=1}^4 e^{2 \pi i (m-k)n /d} \alpha_j^\star(k) \alpha_l(m) \\  &\quad\times  v_j^\dagger(k) v_l(m)  \left[\lambda_j(k)^\star \lambda_l(m)\right]^t. 
\end{split}
\end{equation}

\subsection{Time-averaged limiting probability distribution}\label{sec:limiting-dist}
Following \cite{aharonov01walks}, let us now consider time-averaged probability
distribution
$
\bar{p}(n) = \lim_{t\rightarrow \infty} \frac{1}{t} \sum_{s=0}^{t-1} p(n,s).
$

Using the expression (\ref{eqn:prob-n-t}) for $p(n,t)$ and the fact that the sums
are finite, we get
\begin{equation}\label{eqn:time-aver-prob}
\bar{p}(n) = \sum_{k,m=0}^{d-1}\sum_{j,l=1}^4 K_n(k,j,m,l) \lim_{t\rightarrow\infty}\frac{1}{t}\sum_{s=0}^{t-1}\left[\lambda_{j}^\star(k)\lambda_{l}(m)\right]^s
\end{equation}
where
\begin{equation}
K_n(k,j,m,l) \equiv \frac{1}{d^2}\alpha_{j}^\star(k)\alpha_{l}(m)v_{j}^\dagger(k)v_{l}(m)e^{2 \pi i (m-k)n/d}.
\end{equation}

\begin{figure}[ht!]
    \centering
    
    \subfigure[$d=3$]{
    \includegraphics[width=.45\textwidth]{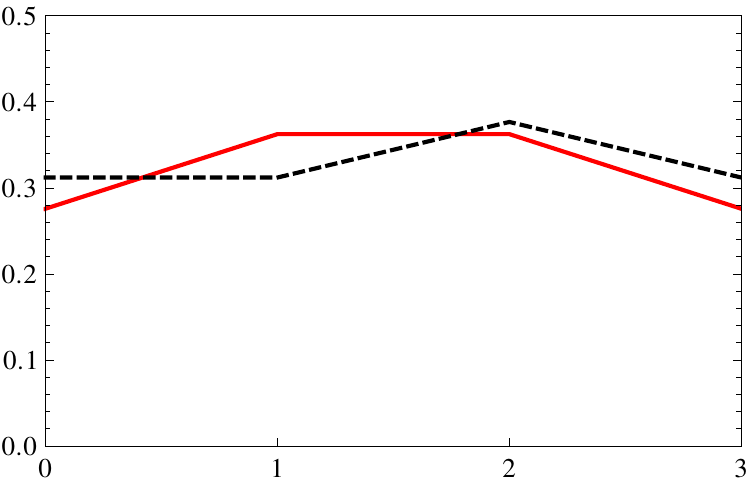}
	\label{fig:lpd-3}
    }
    \subfigure[$d=4$]{
    \includegraphics[width=.45\columnwidth]{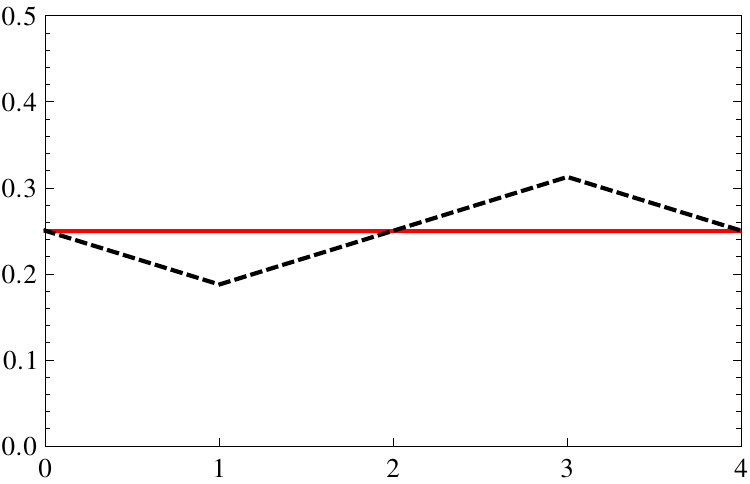}
	\label{fig:lpd-4}
    }\\
    
    \subfigure[$d=6$]{
    \includegraphics[width=.45\columnwidth]{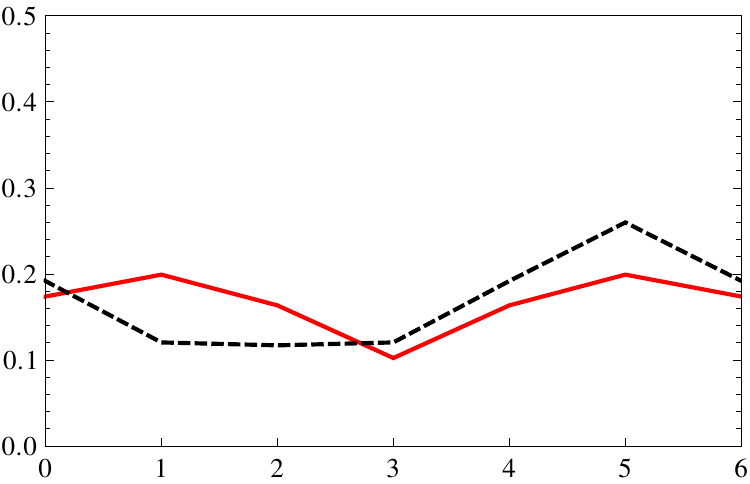}
	\label{fig:lpd-6}
    }
    \subfigure[$d=11$]{
    \includegraphics[width=.45\columnwidth]{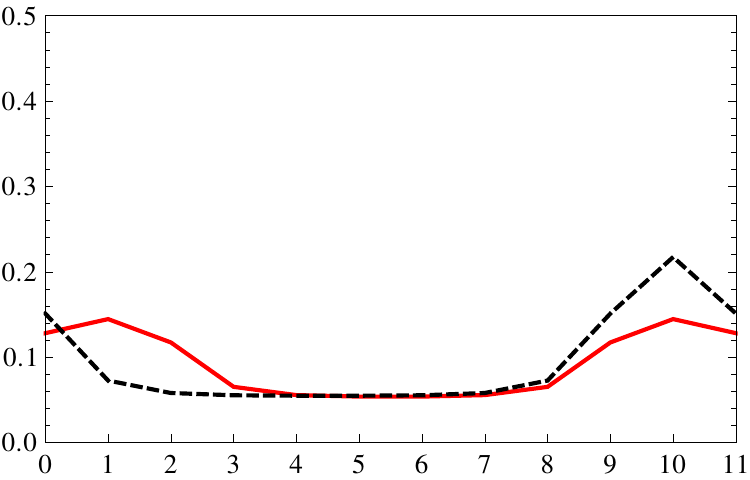}
	\label{fig:lpd-11}
    }
%

    \caption{Limiting probability distributions for Hadamard quantum walks with
    one-step memory on cycle with $d=3,4,6$ and $11$ nodes for the initial
    states given by Eq.~(\ref{eqn:init-state}) (solid red line) and
    $\ket{\phi_1} = (1, 0, 0, 0)^T$ (dashed black line). For each $d$ the
    results are plotted in range $[0,\dots,d]$ to illustrate the periodicity of
    the limiting distribution. Nodes are numbered starting from 0.}
    \label{fig:lpd-examples}
\end{figure}

One can observe that the convergence of $\bar{p}(n)$ depends only on the
behavior of the term
\begin{equation}
f(k,j,m,l) \equiv \lim_{t\rightarrow\infty}\frac{1}{t}\sum_{s=0}^{t-1}\left[\lambda_{j}^\star(k)\lambda_{l}(m)\right]^s.
\end{equation}
The value of this functions depends on the product of eigenvalues as
\begin{equation}\label{eqn:sum-coef-by-lambdas}
f(k,j,m,l) = \left\{
\begin{array}{cr}
1&\ \mathrm{if}\ \lambda_{j}^\star(k)\lambda_{l}(m)=1 \\
0&\ \mathrm{otherwise}\hfill
\end{array}\right..
\end{equation}

Unfortunately, any further simplifications of Eqs.~(\ref{eqn:prob-n-t}) and
\ref{eqn:time-aver-prob}) were not possible. In particular, this simplification
requires a closed form for the product of eigenvalues,
$\lambda_{j}^\star(k)\lambda_{l}(m)$. However, definition in
Eq.~(\ref{eqn:sum-coef-by-lambdas}) can be easily calculated using the standard
computer algebra systems, and thus allows for the evaluation of
Eq.~(\ref{eqn:time-aver-prob}).

\begin{figure}[ht!]
    \centering
    
     \subfigure[$d=200$, $\ket{\phi_0}$]{
    \includegraphics[width=.45\textwidth]{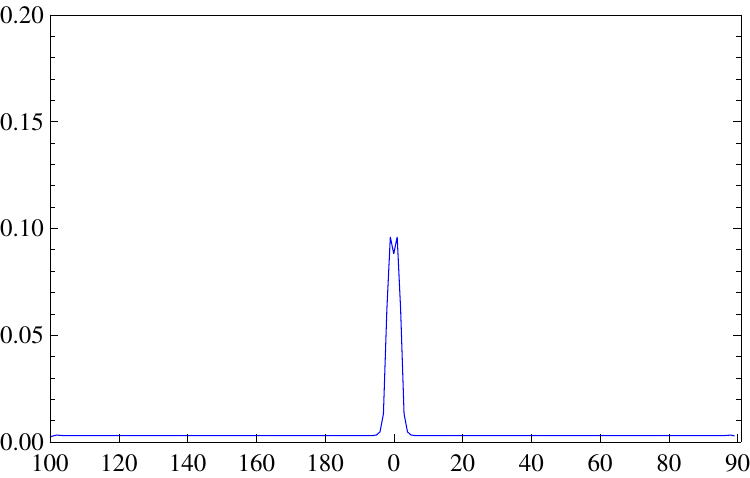}
	\label{fig:lpd-200-0}
    }
    \subfigure[$d=200$, $\ket{\phi_1}$]{
    \includegraphics[width=.45\columnwidth]{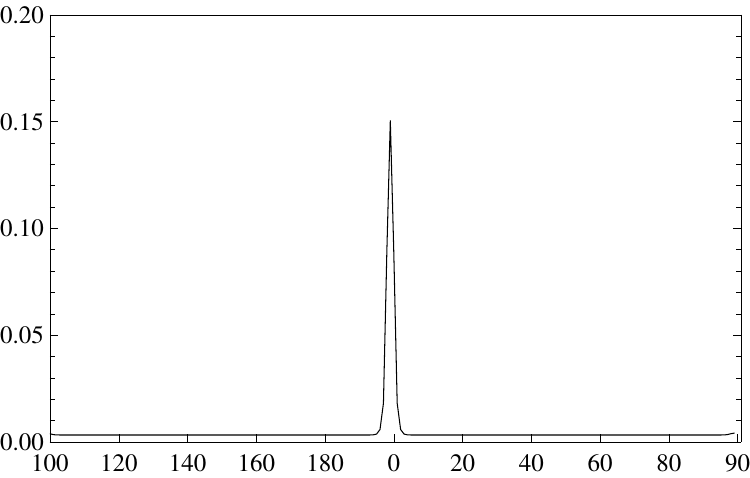}
	\label{fig:lpd-200-1}
    }\\
    
    \subfigure[$d=300$, $\ket{\phi_0}$]{
    \includegraphics[width=.45\columnwidth]{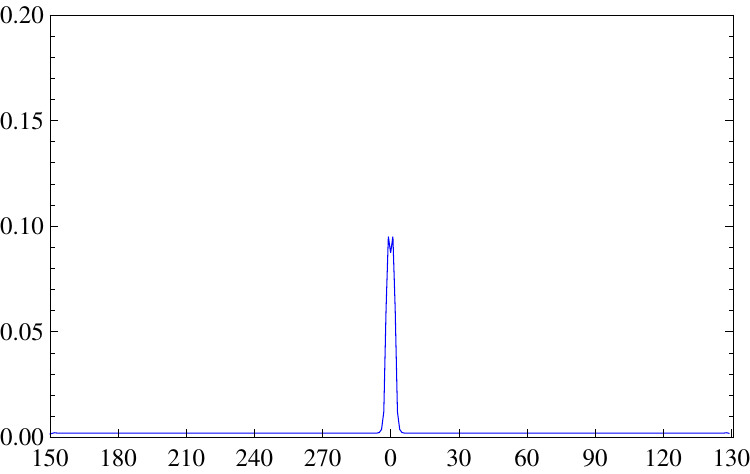}
	\label{fig:lpd-300-0}
    }
    \subfigure[$d=300$, $\ket{\phi_1}$]{
    \includegraphics[width=.45\columnwidth]{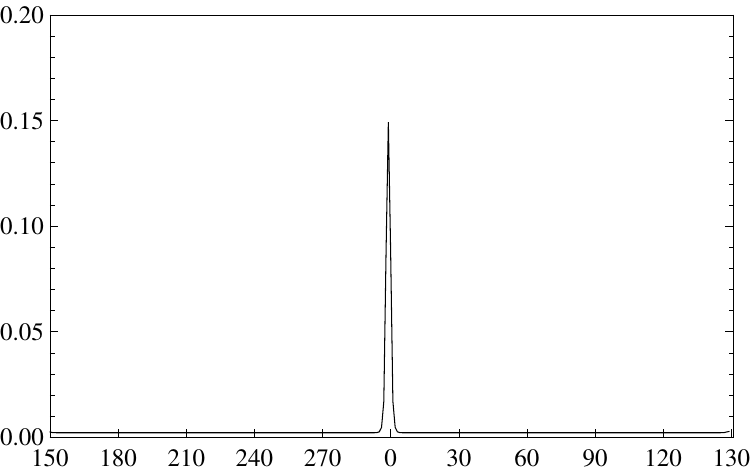}
	\label{fig:lpd-300-1}
    }
    \caption{Limiting probability distributions for Hadamard quantum walks with
    one-step memory on cycle with $d=200$ (\subref{fig:lpd-200-0} and
    \subref{fig:lpd-200-1}) and $d=300$ (\subref{fig:lpd-300-0} and
    \subref{fig:lpd-300-1}) nodes. Result were obtained for initial state
    $\ket{\phi_0}$ -- \subref{fig:lpd-200-0} and \subref{fig:lpd-300-0} -- and
    initial state $\ket{\phi_1}$ -- \subref{fig:lpd-200-1} and
    \subref{fig:lpd-300-1}. For the large number of nodes the only significant
    contributions stem from the nodes close to the starting node.}
    \label{fig:lpd-examples-larger}
\end{figure}

The examples of time-averaged limiting distribution for the discussed model
calculated using Eq.~(\ref{eqn:time-aver-prob}) are presented in
Fig.~\ref{fig:lpd-examples} and Fig.~\ref{fig:lpd-examples-larger}

In order to illustrate the influence of the initial
state on the resulting distribution, two initial states were used to obtain
these plots. In the case of the input state given by Eq.~(\ref{eqn:init-state}),
the memory register is in the superposition, and the resulting time-averaged
probability distribution is symmetric with respect to the starting position. On
the other hand, if the memory register is set to $\ket{0}$ in the initial state,
the resulting time-averaged probability distribution does not have this
property.

Comparison of the limiting probability distribution for larger numbers of nodes
is presented in Fig.~\ref{fig:lpd-examples-larger}. One can see that in these
case the only significant contributions stem from the nodes close to the
starting node.

The important difference in comparison with the memoryless quantum walk on cycle
is that the time averaged limiting distribution depends on the initial state of
the coin and the memory registers, but not on the parity of the number of nodes
$d$~\cite{travaglione02implementing,bednarska03walks}. The dependency is
expressed in the $\alpha_j(k)$ coefficients, which are defined in
Eq.~(\ref{eqn:fourier-init-state-alphas}).

\section{Summary}\label{sec:final-remarks}
We have introduced a model of quantum walk with memory on cycle and studied its
basic properties. We have calculated the probability of finding the particle at
each position after given number of steps and we have provided a formula for the
time-averaged limiting probability distribution for the discussed model. We have
also pointed out the most important differences between quantum walks on cycles
with and without memory. The most important difference is that the symmetry of
the time-averaged limiting probability distribution is independent of the parity
of the number of nodes. However, this distribution is heavily influenced by the
initial state of the memory register.

\section*{Acknowledgments}
Authors would like to thank Claas R\"over and Mike Batty for interesting and
motivating discussions during the preparation of this manuscript. JAM would like
to acknowledge support by the Polish National Science Centre under the research
project UMO-2011/03/D/ST6/00413 and thank de Br\'un Centre for Computational
Algebra, NUI Galway for the hospitality during two research visits.


\bibliographystyle{plain}
\bibliography{qwalks_memory}

\end{document}